\documentclass[
aps,
prx,
reprint,
superscriptaddress
]{revtex4-2}

\usepackage{soul}
\usepackage{amsmath}
\usepackage{comment}
\usepackage{graphicx}
\usepackage{dcolumn}
\usepackage{bm}
\usepackage{hyperref}
\hypersetup{
    colorlinks=false,
    hidelinks
}

\begin{document}

\title{Many-body Interactions in Ultrafast Spectroscopy}

\author{Diego Florio}
\affiliation{Dipartimento di Fisica, Politecnico di Milano, Piazza Leonardo da Vinci 32, 20133 Milano, Italy}
\affiliation{Istituto di Fotonica e Nanotecnologie, Consiglio Nazionale delle Ricerche, Piazza Leonardo da Vinci 32, 20133 Milano, Italy}
\author{Alessandro Baserga}
\affiliation{Dipartimento di Fisica, Politecnico di Milano, Piazza Leonardo da Vinci 32, 20133 Milano, Italy}
\author{Sofia Bertuetti}
\affiliation{Dipartimento di Fisica, Politecnico di Milano, Piazza Leonardo da Vinci 32, 20133 Milano, Italy}
\author{Federico Visentin}
\affiliation{Istituto di Fotonica e Nanotecnologie, Consiglio Nazionale delle Ricerche, Piazza Leonardo da Vinci 32, 20133 Milano, Italy}
\author{Adar Levi}
\affiliation{Institute of Chemistry and The Center for Nanoscience and Nanotechnology, The Hebrew University of Jerusalem, 9190401 Jerusalem, Israel}
\author{Uri Banin}
\affiliation{Institute of Chemistry and The Center for Nanoscience and Nanotechnology, The Hebrew University of Jerusalem, 9190401 Jerusalem, Israel}
\author{Giulio Cerullo}
\affiliation{Dipartimento di Fisica, Politecnico di Milano, Piazza Leonardo da Vinci 32, 20133 Milano, Italy}
\affiliation{Istituto di Fotonica e Nanotecnologie, Consiglio Nazionale delle Ricerche, Piazza Leonardo da Vinci 32, 20133 Milano, Italy}
\author{Franco V. A. Camargo}
\email{franco.camargo@cnr.it}
\affiliation{Istituto di Fotonica e Nanotecnologie, Consiglio Nazionale delle Ricerche, Piazza Leonardo da Vinci 32, 20133 Milano, Italy}

\date{\today}

\begin{abstract}
Nonlinear spectroscopy is widely used to probe ultrafast dynamics in matter. Signals of higher nonlinear orders encode valuable many-body dynamics but only recently have become individually retrievable. However, their physical interpretation in terms of double-sided Feynman diagrams is cumbersome and non-intuitive. Here, we establish an alternative framework that connects them to the underlying excitation conditions and many-body interactions. This approach predicts two classes of dynamic nonlinearities, due either to direct many-body interactions or to saturable kinetic channels. The method provides a direct physical interpretation of the retrieved signals, explaining their sign, magnitude, and dynamics. It also enables model-free reconstruction of the dynamics of a  sub-ensemble of particles starting with a given number of excitons, which we demonstrate to be especially insightful. We validate the framework using colloidal CdSe/CdS core/shell quantum dots, which enable us to model nonlinearities of both classes in isolation and also combined.
\end{abstract}

\maketitle

\section{Introduction}
Many-body dynamics play a central role in the physics of photoexcited systems, including important phenomena such as super-radiance \cite{Dicke1954, Rainò2018}, amplified spontaneous emission \cite{Yakunin2015, Kobiyama2026}, Auger recombination \cite{Klimov2000, Klimov2014}, state filling \cite{Ghosh2019, Tekelenburg2024}, and hot-phonon bottlenecks \cite{Yang2015, Fu2023}. Time-resolved spectroscopies and, in particular, pump-probe spectroscopy provide means to investigate these dynamics down to ultrafast timescales. Normally, pump-probe spectroscopy is performed in the low-excitation, perturbative, regime, in which the signal scales linearly with the excitation intensity $I$. By photoexciting at higher intensities, one can progressively increase the contribution of nonlinear processes to the total response. In practice, in the monomolecular regime one obtains a 3$^\mathrm{rd}$-order nonlinear optical signal due to two light-matter interactions with the pump and one with the probe field, which scales linearly with $I$. At higher $I$, nonlinear phenomena such as multiphoton excitation, absorption saturation, and many-body kinetics lead to deviations from linearity, and the signal can be expressed as a Taylor expansion:
\begin{equation}
S(I)=\sum_{n=1}^\infty S^{(2n+1)}I^n=S^{(3)} I+S^{(5)} I^2+S^{(7)} I^3+\dots,
\label{Intensity_expansion}
\end{equation}
where $S$ is the measured signal and $S^{(2n+1)}$ is the contribution of the $\mathrm{n^{th}}$ nonlinear order, where generally only odd orders contribute due to material symmetry. However, in practice intensity-dependent datasets were mostly analyzed phenomenologically until \cite{Righetto2018, Valkunas1996, Soavi2016, Bressan2020, Lenngren2013}, recently, Malý~\textit{et al.} proposed a method to invert Eq.~\ref{Intensity_expansion} and retrieve $S^{(2n+1)}$ directly from experimental data \cite{Maly2023, Luttig2023_JCP}. In the context of many-body interactions, retrieving the individual contributions $S^{(2n+1)}$ explicitly is very appealing, since their $I^n$ scaling suggests a direct connection to dynamics involving $n$ excitations. Since nonlinearities arising from the direct interaction with the external fields, i.e., absorption saturation or multi-photon excitation, are present from time zero, many-body interactions can be retrieved from the dynamic evolution of the higher order $S^{(2n+1)}$ signals, usually described in terms of large numbers of double-sided Feynman diagrams \cite{Dostal2018,Rose2021,Maly2023,Rose2023}. Although these diagrams have proven extremely successful in interpreting nonlinear spectroscopy signals in general \cite{Nemeth2010,Meneghin2018, Tollerud2014, Lim2015, Thyrhaug2016, Policht2018, Camargo2017, Abramavicius2020}, they are often not intuitive and directly helpful to experimentalists aiming to understand relevant many-body interactions present in a given dataset.

Here, we take an entirely different approach to the description of many-body nonlinearities in transient absorption signals. Starting from Poisson excitation statistics, we derive relations between the nonlinear signals $S^{(2n+1)}$ and the much more intuitive dynamics of a sub-ensemble of particles that have undergone exactly $n$ excitations, which we call $d_n$. The relation obtained is purely statistical and readily interpretable. We also derive a recursive set of equations to connect $d_n$ to the underlying many-body interaction mechanisms, from which we identify two classes of dynamic nonlinearities. The most often considered is that of explicit interactions between different excitations, such as Auger or exciton-exciton annihilation processes. The second one concerns dynamic saturation due to competition for states of limited degeneracy, such as state filling. Its value lies in the fact that, for any given set of many-body interactions, the dynamics of a sub-ensemble starting with $n$ excitations are easy to visualize. By contrast,  the physical interpretation of $S^{(2n+1)}$ is far less immediate. Indeed, we separately consider two nonlinearities, one from each class, and provide the physical interpretation of the resulting nonlinear signals by combining the insight from these equations with the intuitive nature of $d_n$. Finally, we consider, a case study in which the two classes of nonlinearities are simultaneously present. Although this produces confusing $S^{(2n+1)}$ features, the interpretation of $d_n$ remains possible. The first and last models are supported by experimental data on colloidal CdSe/CdS core-shell quantum dots (QDs), while the method itself is general.

\section{Results}
\subsection{Connecting high-order nonlinear signals to many-body interactions}

\begin{figure}[hbtp]
\centering
\includegraphics[scale=1]{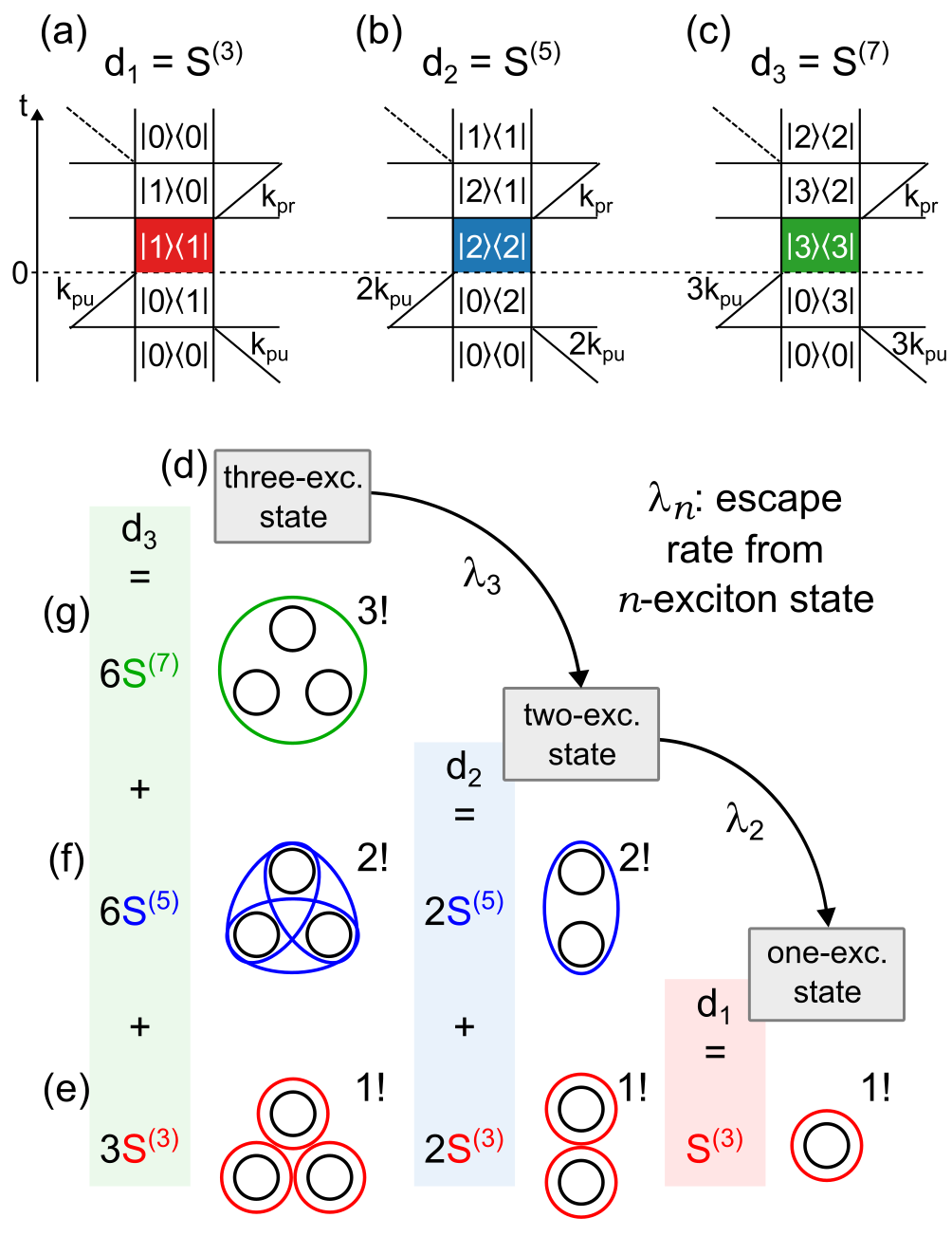}
\caption{(a--c) Typical double-sided Feynman diagrams corresponding to one exciton (a), two exciton (b), and (c) three exciton populations. (d) Markovian population kinetics from the three-exciton state down to the one-exciton state. $\lambda_n$ is the escape rate from the $n$-exciton state. (e--g) Combinatorial origin of the coefficients entering the nonlinear-order reconstruction of the population dynamics initialized with one, two, and three excitons.}
\label{fig1}
\end{figure}

Let us first consider an ensemble of absorbing nano-objects with a high density of states in resonance with the pump pulses, so that within this ensemble absorption events obey Poisson statistics. The ensemble can then be partitioned into sub-ensembles of nano-objects excited once, twice, etc., by the pump pulse. Inspection of the double-sided Feynman diagrams in Fig. \ref{fig1}(a-c) may tempt one to assign them, and therefore $S^{(2n+1)}(t)$, directly to the dynamics $d_n(t)$ of the sub-ensemble containing $n$ excitations. However, the $n$ excitations present within a nano-object may or may not interact; in the absence of interaction, the measured signal is simply $n$ times the third-order signal. Hence, although diagrams such as those in Fig. \ref{fig1}(b,c) appear to strongly imply a one-to-one correspondence between the dynamics $d_n(t)$ of a sub-ensemble of particles with $n$ excitations and $S^{(2n+1)}(t)$, this is not generally the case \cite{Rose2021,Rose2023,Krich2025}. In the following paragraphs, we establish a connection between these two quantities and show how the physical meaning of $S^{(2n+1)}(t)$ emerges naturally.

To derive the relationship between $d_n(t)$ and $S^{(2n+1)}(t)$, we assume that the number $n$ of excitations initially generated in a particle follows a Poisson distribution:
\begin{equation}
P_N(n)=\frac{N^{n}}{n!}e^{-N},
\label{poisson}
\end{equation}
where $N$, the average number of excitations per particle, is dependent on the pump intensity. Assuming that the signal is proportional to the total number of excitations, the total signal as a function of $d_n(t)$ can be written as:
\begin{equation}
S(I,t)=\sum_{n=1}^{\infty}P_N(n)\,d_n(t).
\label{average_is_the_number}
\end{equation}
Although $d_n(t)$ are not generally accessible experimentally, they are physically intuitive quantities, since they describe how the presence of additional excitations modifies the subsequent dynamics. In the absence of many-body effects, one would simply have $d_n(t)=n\,d_1(t)$, whereas deviations from this linear scaling contain information on many-body kinetics. To establish an explicit connection between $d_n(t)$ and $S^{(2n+1)}$, we substitute the Poisson distribution of Eq.~\ref{poisson} into Eq.~\ref{average_is_the_number}, expand the resulting ensemble response in powers of $N$, and identify the coefficient of each power $N^n$ with the corresponding nonlinear order (see Supplementary Note \ref{SI_derivation_dn_to_nls}). This gives
\begin{equation}
d_n(t)=\sum_{k=1}^{n}\binom{n}{k}k!\,S^{(2k+1)}(t).
\label{dn_to_S(2k+1)}
\end{equation}
Eq.~\ref{dn_to_S(2k+1)} shows that the $d_n(t)$ are not, in general, associated with a single nonlinear order, as Fig.~\ref{fig1}(a--c) might suggest. Rather, the mapping between $d_n(t)$ and the nonlinear components is one-to-many in the form of a triangular system. Thus, we introduce a distinct index $k=1,\dots, n$ to label the nonlinear signals hereafter. This triangular mapping reveals that the $(2k+1)$-order component $S^{(2k+1)}(t)$ isolates the additional dynamics that first emerge at the $k$-excitation level and contribute to all dynamics $d_n(t)$ with $n\geq k$.  In this sense, $S^{(2k+1)}(t)$ does encode correlations among $k$ excitations, as expected. Moreover, the combinatorial coefficients weighting the nonlinear signals in Eq.~\ref{dn_to_S(2k+1)} have a clear physical meaning. The binomial factor $\binom{n}{k}$ counts the number of distinct $k$-excitation subsets that can be formed from the $n$ excitations present at time zero, and the factor $k!$ accounts for the permutation multiplicity associated with the $k$ independent excitation events corresponding to each subset. Figures~\ref{fig1}(e--g) help visualize Eq.~\ref{dn_to_S(2k+1)} with $1 \le n \le3$. Each column displays $d_n(t)$ as given by Eq. \ref{dn_to_S(2k+1)}, while each row shows how the contribution of each nonlinear signal $S^{(2k+1)}(t)$ to the sub-ensemble dynamics changes as more excitations are added. Reading each column gives the reconstruction of a specific sub-ensemble dynamics: $d_1(t)$ contains only $S^{(3)}(t)$, $d_2(t)$ contains $S^{(3)}(t)$ and $S^{(5)}(t)$, and $d_3(t)$ additionally requires $S^{(7)}(t)$, each with the combinatorial weights prescribed by Eq.~\ref{dn_to_S(2k+1)}. Conversely, reading Fig. \ref{fig1}(e-g) by rows shows how a given nonlinear order contributes across excitation-number sectors: $S^{(3)}(t)$ enters every $d_n(t)$, whereas progressively higher orders first emerge only when the corresponding excitation number is reached. 

Crucially, this reconstruction is model-independent. Since the nonlinear signals $S^{(2k+1)}(t)$ are now experimentally accessible via the method of Malý~\textit{et al.}, Eq. \ref{dn_to_S(2k+1)} provides a direct, model-free route to the more intuitive quantities $d_n(t)$, as we demonstrate below. The purely statistical connection between $d_n(t)$ and $S^{(2k+1)}(t)$, however, raises the question: what determines the underlying dynamics? The  $d_n(t)$ are naturally connected to the usual master-equation description of population kinetics \cite{Gillespie1977, Barzykin2007}, from which the following recursive decomposition can be derived (see Supplementary Note \ref{SI_derivation_recursive_dn}):
\begin{equation}
d_n(t)
=
n\,e^{-\lambda_n t}
+
\int_0^t
\lambda_n
\,e^{-\lambda_n\tau}
\,\widetilde d_{n-1}(t-\tau)
\,\mathrm{d}\tau ,
\label{convolution}
\end{equation}
where $\lambda_n$ is the total rate for leaving the $n$-excitation state through the $n\rightarrow n-1$ transition, as indicated by the arrows connecting successive excitation-number sectors in Fig.~\ref{fig1}(d) and $\widetilde d_{n-1}$ denotes the sub-ensemble dynamics of particles with $n-1$ excitations prepared via relaxation from $d_n$. Eq.~\ref{convolution} simply separates the population dynamics according to whether a kinetic event has occurred. The first term describes trajectories for which no event has occurred up to time $t$: the system therefore still contains all $n$ initial excitations, with probability $e^{-\lambda_n t}$. The second term describes trajectories in which the first event occurs at an earlier time $\tau$. The factor $\lambda_n e^{-\lambda_n\tau}\mathrm{d}\tau$ gives the probability that this first event occurs between $\tau$ and $\tau+\mathrm{d}\tau$; after the event, the system evolves from the state reached for the remaining time $t-\tau$, according to $\widetilde d_{n-1}(t-\tau)$. The integral therefore averages over all possible times at which the first event can occur and over the ensuing dynamics. Eq.~\ref{convolution} can be straightforwardly generalized to transitions with different step sizes or to multiple competing kinetic channels by introducing the corresponding event rates and conditional post-transition dynamics. 

Importantly, Eq.~\ref{convolution} identifies two distinct classes of dynamic nonlinearities. The most commonly considered case arises from direct many-body interactions, such as Auger recombination or exciton-exciton annihilation, for which the escape rate $\lambda_n$ explicitly acquires a nonlinear dependence on the excitation number through the interaction itself. Interestingly, however, Eq.~\ref{convolution} also reveals a second source of nonlinearity: an event that alters the kinetic channels available to the remaining population, so that $\widetilde d_{n-1}(t)\neq d_{n-1}(t)$. The most common example is state-filling, in which many excitations compete for a state whose degeneracy is smaller than the total population. This competition imposes a correlation on the dynamics of all excitations which failed to occupy this state while still available \cite{Ghosh2019, De2025}.

\begin{figure*}[hbtp]
\centering
\includegraphics[scale=1]{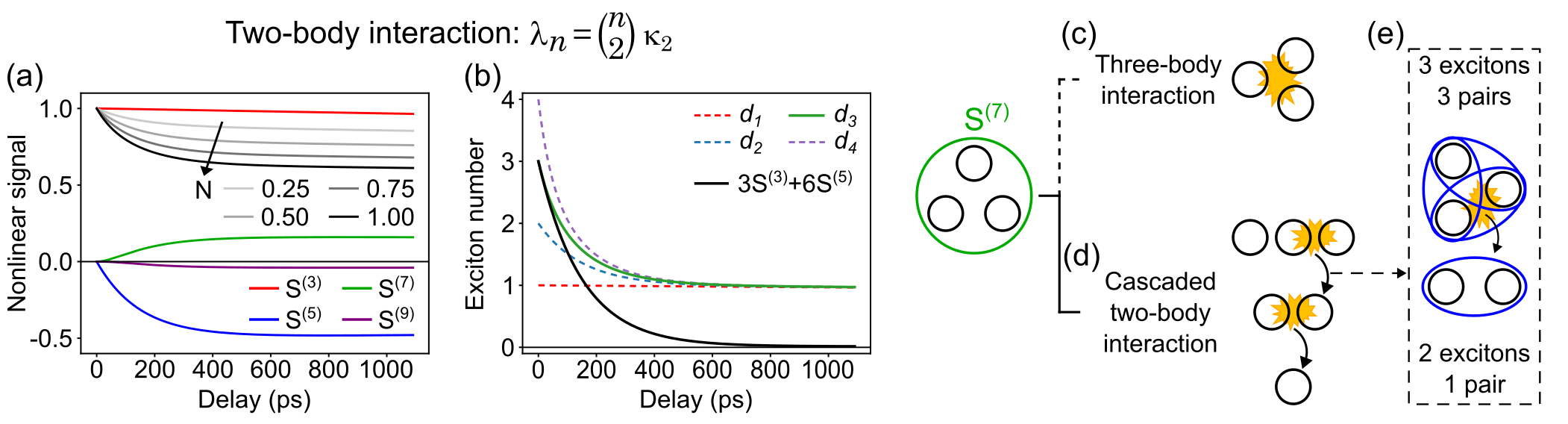}
\caption{(a) Normalized simulated intensity-dependent dynamics for values of $N$ of $0.25$, $0.5$, $0.75$, $1$ for the model with the escape rate in Eq.~\ref{pairwise_escape_rate} are shown in grayscale. 3$^{\mathrm{rd}}$-, 5$^{\mathrm{th}}$-, 7$^{\mathrm{th}}$-, and 9$^{\mathrm{th}}$-order signals  are shown in color. (b) Dynamics for a sub-ensemble of particles with 1 to 4 excitons at $t=0$. The partial reconstruction of the three-exciton dynamics $3S^{(3)}+6S^{(5)}$ is shown in black. (c--e) Possible origins of $S^{(7)}$. (c) Direct three-body interaction. (d) Successive pairwise interactions along the $3\rightarrow2\rightarrow1$ cascade. (e) Number of pairs before and after the first recombination event.}
\label{fig2}
\end{figure*}

\subsection{Interaction-driven dynamic nonlinearities}
\label{sec:interaction-driven_nl}

We now turn to the physical interpretation of the nonlinear signals $S^{(2k+1)}(t)$ for both classes of dynamic nonlinearity outlined above. We first consider two models, each including a single nonlinear process corresponding to one of the classes identified in Eq.~\ref{convolution}, and then address the case in which both are simultaneously present. The first model combines single-body recombination, analogous to radiative recombination of an exciton, with exciton-exciton annihilation, a widespread two-body recombination mechanism~\cite{Daniel2007, Kriete2019, Maly2020, Deng2020, Luttig2021, Maly2023, Luttig2023_JPCL, Kumar2023, Zheng2024, Florio2025, Bruschi2026, Sokolovskii2026, Fairbairn2026, Braun2026}. Both processes take the system from $n$ to $n-1$ excitons, so the state reached after either event is kinetically equivalent to one originally prepared with $n-1$ excitons. Hence, $\widetilde d_{n-1}(t)=d_{n-1}(t)$, and no saturation-related nonlinearities can arise. In this model, the escape rate of the $n$ exciton state is:
\begin{equation}
\lambda_n
=
\kappa_1\,n
+
\kappa_2\binom{n}{2},
\label{pairwise_escape_rate}
\end{equation}
where the single-body process with rate $\kappa_1$ scales linearly with $n$, while the pairwise process, with rate $\kappa_2$ scales with the number of distinct exciton pairs $\binom{n}{2}$, introducing a quadratic dependence on $n$. This case is physically interesting not only because it is ubiquitous, but also because it includes both a 3$^\mathrm{rd}$- and a 5$^\mathrm{th}$-order relaxation mechanism. 

The gray lines in Fig.~\ref{fig2}(a) show normalized intensity-dependent traces simulated with the model in Eq.~\ref{pairwise_escape_rate} for different values of the average excitation number $N$ over a 1 ns observation window, where radiative recombination is negligible. The chosen rates,  $\kappa_1$ = $(30\,\mathrm{ns})^{-1}$ and  $\kappa_2$ = $(150\,\mathrm{ps})^{-1}$, are based on experimental data on QDs shown in Supplementary Note \ref{SI_experimental_data}. The dynamics exhibit progressively stronger depletion with increasing $N$. This behavior reflects the growing contribution of multiexciton states undergoing exciton-exciton annihilation, whose rate accelerates at higher $N$ because the number of available exciton pairs increases as $\binom{n}{2}$. The corresponding nonlinear contributions $S^{(2k+1)}(t)$, which capture these intensity-dependent dynamics, are shown in color in Fig.~\ref{fig2}(a). The results are consistent with those reported in exciton-exciton annihilation studies of other systems \cite{Maly2023, Dietzsch2023, Lee2024, Shi2025, Zhang2025, Lee2025, Buttner2025, Binzer2026, Sebelik2026, Keil2026, Braun2026}. $S^{(3)}(t)$, which is expected to contain only the monomolecular response, shows very little decay over the 1 ns time window, confirming that it is governed solely by radiative recombination. In contrast, $S^{(5)}(t)$ is negative and increases in magnitude on the 150 ps exciton-exciton annihilation timescale, as expected for a two-body mechanism which is intrinsically fifth order. The negative sign is intuitively explained by the fact that exciton-exciton annihilation causes a reduction of the overall positive signal.  We must now interpret the physical origin of $S^{(7)}(t)$ and $S^{(9)}(t)$, which display alternating signs and progressively delayed rise times, despite the absence of any direct seventh or ninth-order process in this system.

To address this question, we consider the sub-ensemble dynamics of particles initially containing one to four excitons, reconstructed from $S^{(2k+1)}(t)$ via Eq.~\ref{dn_to_S(2k+1)}, shown in dashed lines in Fig.~\ref{fig2}(b). Even though it is immediately clear that $S^{(2k+1)}(t)$ cannot be identified one-to-one with $d_n(t)$ with $n=k$, the latter have a direct physical interpretation. While $d_1(t)$ starts from one exciton, which can thus only recombine radiatively on the 30 ns timescale, exciton-exciton annihilation acts progressively faster on $d_{2-4}(t)$ bringing the population from its initial value to one exciton, after which radiative recombination is the only available channel. We can now clarify the physical meaning of $S^{(7)}(t)$ by analyzing the kinetic role of each nonlinear signal in the reconstruction of $d_3(t)$, which from Eq. \ref{dn_to_S(2k+1)} is:

\begin{equation}
d_3(t)=3 S^{(3)}(t) + 6 S^{(5)}(t) + 6 S^{(7)}(t).
\label{d3}
\end{equation}

The first term clearly describes the case in which the three excitons evolve independently; their response is then simply the sum of three single-exciton responses.  The second term accounts for exciton-exciton annihilation. Its statistical weight $\binom{3}{2}2!=6$ arises from the $\binom{3}{2}=3$ distinct exciton pairs and the $2!$ permutations of the two independent excitation events that generated them. The role of the third term is much more interesting, since in this system there is no direct many-body process connecting three excitons. To understand the underlying physics, it is instructive to consider the partial reconstruction given by Eq. \ref{d3} without including the last term, i.e., $3S^{(3)}(t)+6S^{(5)}(t)$, which is shown in black in Fig. \ref{fig2}, and compare it with $d_3(t)$ (green line). The two curves agree well at first, but as the exciton population drops from 3 to 2 they diverge strongly. The partial reconstruction incorrectly brings the population down to 0 instead of 1 after the annihilation timescale. This occurs because, after the first recombination event, the number of exciton pairs abruptly decreases from $\binom{3}{2}2!=6$ to $\binom{2}{2}2!=2$ (Fig. \ref{fig2}(e)). Therefore, the coefficient of $S^{(5)}(t)$, that correctly accounts for the faster annihilation kinetics of the three-exciton state, needs to be updated after the first exciton is lost. This is accomplished by the $S^{(7)}(t)$ correction term, which can be physically understood as the additional dynamics generated by successive pairwise interactions along the $3\rightarrow2\rightarrow1$ cascade shown in Fig.~\ref{fig2}(d). Both the sign of $S^{(7)}(t)$ and its delayed buildup then become clear: the correction becomes relevant only as the population reaches the two-exciton state through the preceding interaction, and it prevents over-depletion of the population caused by an $S^{(5)}(t)$ coefficient that no longer applies. Its weight in $d_3(t)$ is $\binom{3}{3}3!=6$, corresponding to the single three-excitation subset and the $3!$ permutations of its independent excitation events, as illustrated in Fig.~\ref{fig1}(g). For larger initial excitation numbers, progressively higher nonlinear orders are required to account for the changing number of pairwise interaction channels along longer cascades. Accordingly, $S^{(9)}(t)$ in Fig.~\ref{fig2}(b) captures the additional contribution associated with the $4\rightarrow3\rightarrow2\rightarrow1$ sequence, and is further delayed because two recombination events are needed to create a correlation among four excitons.

\begin{figure*}[hbtp]
\centering
\includegraphics[scale=1]{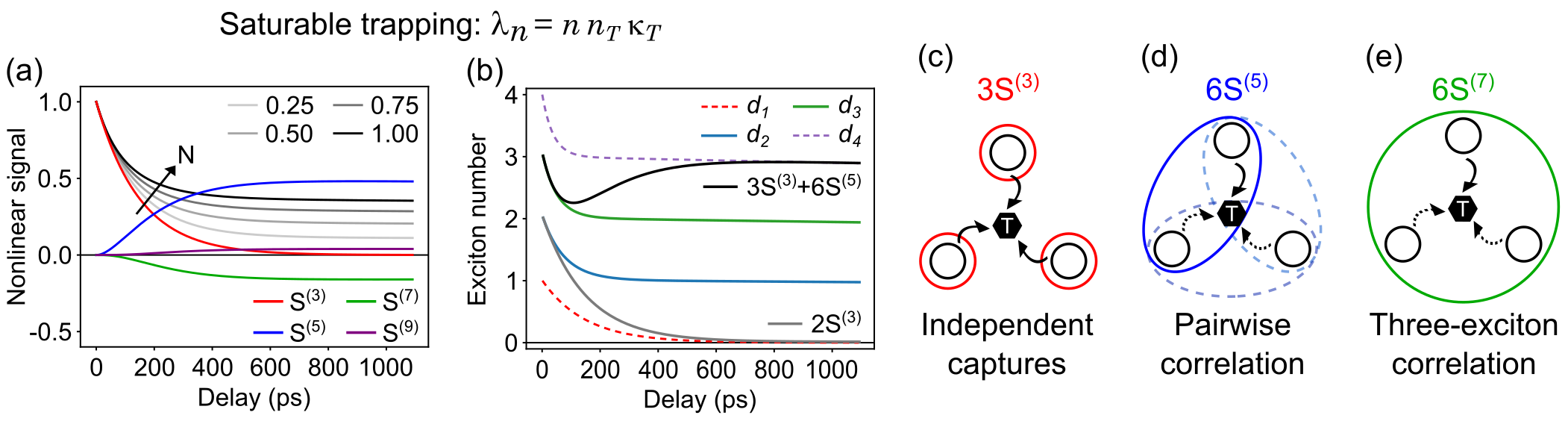}
\caption{(a) Normalized simulated intensity-dependent dynamics for values of $N$ of $0.25$,$0.5$, $0.75$, $1$, for the model with the escape rate in Eq.~\ref{trapping_escape_rate} are shown in grayscale. 3$^{\mathrm{rd}}$-, 5$^{\mathrm{th}}$-, 7$^{\mathrm{th}}$-, and 9$^{\mathrm{th}}$-order signals are shown in color. (b) Dynamics for a sub-ensemble of particles with 1 to 4 excitons at $t=0$. The partial reconstruction of the two-exciton dynamics $2S^{(3)}$ is shown in gray, while the reconstruction of the three-exciton dynamics $3S^{(3)}+6S^{(5)}$ is shown in black. (c--e) Schematic interpretation of the nonlinear hierarchy for a particle initialized with three excitons and a single saturable trapping site. (c) The $3S^{(3)}$ contribution represents independent capture of the three initially present excitons. (d) $6S^{(5)}$ introduces the three possible pairwise competition corrections, shown as overlapping pairwise capture subsets. (e) $6S^{(7)}$ accounts for the three-exciton correlation arising from the common overlap of these pairwise corrections, required because all three excitons compete for the same trap.}
\label{fig3}
\end{figure*}

\subsection{Saturation-driven dynamic nonlinearities}

Now we  turn to the second class of dynamic nonlinearities, whose nonlinear signals have not yet been interpreted in the framework of the method by Malý~\textit{et al.} We consider a kinetic model that includes radiative recombination and carrier trapping at a single defect site, but excludes exciton-exciton annihilation. Crucially, after the first trapping event has occurred, no further trapping is possible, so that in Eq.~\ref{convolution} we have that $\widetilde d_{n-1}(t)\neq d_{n-1}(t)$. This situation is also relevant to QDs \cite{Gigi2025, Gigi2026} where, despite major advances in synthesis and surface passivation that have made near-unity photoluminescence quantum yield routinely accessible \cite{Hanifi2019, Ji2020}, defect-mediated trapping can still become significant, for example as a result of particle aging or degradation. Since the pristine samples are trap-free, it is reasonable to assume that the onset of degradation manifests through the formation of a single trap site in a fraction of the QDs. For physical clarity, we first consider the idealized case in which trapping is present but exciton-exciton annihilation is absent, and then consider the case in which both mechanisms coexist. We describe each particle by the state $(n,n_T)$, where $n$ is the exciton number and $n_T=1$ or $0$ denotes whether the trap state is available or occupied. To isolate this mechanism, we set the defect probability to $p_T=1$, so that every particle initially contains one trap. In addition to the same single-body recombination process considered above, with $\kappa_1=(30\,\mathrm{ns})^{-1}$, each exciton can be trapped with a rate $\kappa_{T}=(150\,\mathrm{ps})^{-1}$. The corresponding transitions are $(n,n_T)\xrightarrow{\,n \kappa_1\,}(n-1,n_T)$, for radiative recombination, and $(n,1)\xrightarrow{\,n \kappa_{T}\,}(n-1,0)$ for trapping. Before the trap is occupied, the total exciton decay rate is:
\begin{equation}
\lambda_{n}=n(\kappa_1+\kappa_{T}),
\label{trapping_escape_rate}
\end{equation}
which remains linear in $n$, implying that both processes contribute to the third-order signal. The many-body nonlinearity does not originate from this excitation-number dependence, but from the change in kinetic state produced by the trapping event: once one exciton is captured, $n_t$ changes from $1$ to $0$ and the trapping channel is no longer available to the remaining population. Consequently, the post-trapping dynamics of the resulting $(n-1,0)$ state differs from the dynamics $d_{n-1}(t)$ of a particle initially prepared in the $(n-1,1)$ state, so that $\widetilde d_{n-1}(t)\neq d_{n-1}(t)$.

The gray lines in Fig.~\ref{fig3}(a) show normalized, intensity-dependent dynamics simulated with the model with the escape rates in Eq.~\ref{trapping_escape_rate} for different values of the average excitation number $N$. The fractional population loss associated with trapping decreases as $N$ increases, reflecting the fact that only one exciton per particle can be trapped. The corresponding nonlinear contributions $S^{(2k+1)}(t)$, associated with these intensity-dependent dynamics, are shown in color in Fig.~\ref{fig3}(a). We start by noticing that $S^{(3)}(t)$ decays to zero on the trapping timescale, as one would expect from Eq. \ref{trapping_escape_rate} and the relative values set between the rates. Interestingly, $S^{(3)}(t)$ and $S^{(5)}(t)$ have the same sign, unlike the case of exciton-exciton annihilation; for higher orders, however, the usual alternating sign pattern is recovered. We also note that higher order nonlinearities once again appear with progressively longer delayed rises. 

To understand these features, we again consider the sub-ensemble dynamics $d_n(t)$, reconstructed from $S^{(2k+1)}(t)$ via Eq.~\ref{dn_to_S(2k+1)} and shown in colored lines in Fig. \ref{fig3}(b). Once again, the $d_n(t)$ have an intuitive interpretation: $d_1(t)$ decays to zero on the trapping timescale, whereas for $n>1$ the trapping becomes gradually faster, because the additional excitons compete for the same state. In all cases, however,  trapping removes only one exciton, and the remaining ones recombine radiatively on a much longer timescale. To break down the connection given by Eq. \ref{dn_to_S(2k+1)}, we note that for a single exciton, $d_1(t)=S^{(3)}(t)$ decays on the characteristic trapping timescale, as expected for a trapping rate that scales linearly with exciton number. For $n=2$, $d_2(t)=2S^{(3)}(t)+2S^{(5)}(t)$. The term $2S^{(3)}(t)$, shown in gray in Fig.~\ref{fig3}(a),  describes two independent copies of the single-exciton dynamics and therefore assumes that both excitons can eventually be captured, overestimating the population loss. The actual $d_2(t)$ initially decays faster than $d_1(t)$, because either exciton can be trapped, but approaches a population of one after the trapping timescale. The positive contribution $2S^{(5)}(t)$ corrects this excess depletion by accounting for the set of trajectories in which both excitons would have been trapped by time $t$ if the trap had remained available after the first trapping event. 

For $n=3$, $d_3(t)$ is once again given by Eq.~\ref{d3}, which is model-independent. By analogy with the $n=2$ case, the term $3S^{(3)}(t)$ corresponds to independent trapping of the three excitons (Fig.~\ref{fig3}(c)) and would therefore drive the reconstructed population from three to zero. The physical role of the higher-order terms can then be read directly from the successive reconstruction in Fig.~\ref{fig3}(b). The positive contribution $6S^{(5)}(t)$, arising from the three possible exciton pairs (Fig.~\ref{fig3}(d)), corrects the unphysical processes in which two excitons are captured by the same trap. Taken together, however, the three pairwise corrections overcompensate the independent-trapping result, producing the nonphysical population revival seen in the partial reconstruction $3S^{(3)}(t)+6S^{(5)}(t)$ and bringing the population back to three, not two excitons. The negative contribution $6S^{(7)}(t)$ provides the remaining correction, reducing the reconstructed population from three to the physically correct value of two. In this picture, $S^{(7)}(t)$ is the three-exciton correlation generated by competition for the same saturable trap (Fig.~\ref{fig3}(e)): it corrects for the fact that the three pairwise $S^{(5)}$ contributions are themselves correlated, because they all involve the same finite-capacity trapping channel.

The same reasoning extends to larger initial excitation numbers: progressively higher nonlinear orders are required to correct for the overcounting that arises when increasing numbers of excitons compete for the same saturable trap. This also explains the temporal hierarchy of the nonlinear signals. Although all excitons compete for the trap from $t=0$, and only one physical trapping event can occur, the corresponding many-exciton corrections build up at different rates. At short times, it is relatively likely that one exciton would have been captured in the hypothetical independent-trapping reference represented by $nS^{(3)}(t)$; it is less likely that two excitons would both have been captured, and less likely still that three excitons would all have been captured within the same time interval. The pairwise competition encoded in $S^{(5)}(t)$ therefore becomes relevant before three-exciton competition encoded in $S^{(7)}(t)$, while correlations involving larger numbers of excitons emerge progressively later. Therefore, although no sequence of additional physical trapping events occurs, higher nonlinear orders still show a delayed buildup. This delay reflects the increasing time required for the independent-trapping reference to generate multiple hypothetical capture events, which must be corrected because the actual system contains only a single trap state.

\begin{figure*}[hbtp]
\centering
\includegraphics[scale=1]{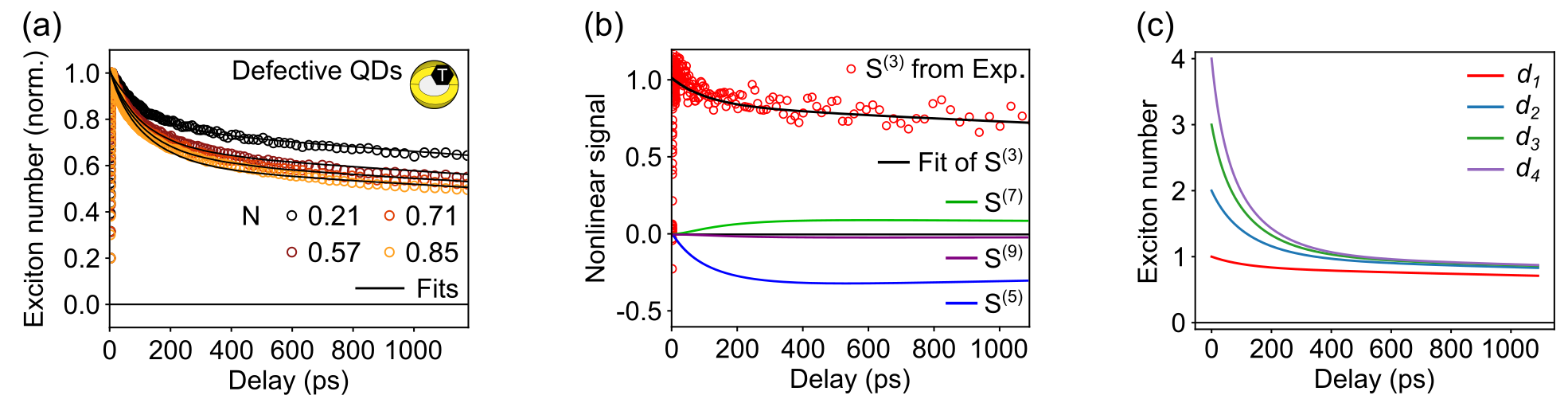}
\caption{(a) Transient absorption dynamics of the band-edge states following 3.1 eV excitation normalized to the average number of excitons per QD, \textit{N}. The fits are shown as black solid lines. (b) 3$^{\mathrm{rd}}$--order component extracted from the fluence-dependent dynamics in panel (a) (red circles). The black solid line shows the fit obtained using a stochastic kinetic model containing radiative recombination and carrier trapping. The 5$^{\mathrm{th}}$-, 7$^{\mathrm{th}}$-, and 9$^{\mathrm{th}}$-order signals, extracted from the fits of the fluence-dependent dynamics of panel (a), are shown as colored solid lines. (c) Dynamics for a sub-ensemble of particles with 1 to 4 excitons at $t=0$.}
\label{fig4}
\end{figure*}

\subsection{Combination of the two classes of nonlinearities}

Having established the nonlinear signatures of a single saturable trapping channel in the idealized limit where every particle contains one defect ($p_T=1$) and exciton-exciton annihilation is absent, we now turn to a realistic ensemble of defective CdSe/CdS core--shell QDs in solution. In contrast to the homogeneous model considered above, the experimental sample is expected to be heterogeneous: only a fraction $p_T$ of the QDs contains one active defect, while the remaining fraction $1-p_T$ is trap-free. This heterogeneity introduces an additional fit parameter compared with the simplified trapping model. A further complication is that, in the real sample, carrier trapping does not act in isolation. Multiexciton states can also undergo exciton-exciton annihilation. Because trapping and annihilation occur on similar timescales~\cite{Smith2015, Das2020}, their respective contribution to the measured dynamics can be difficult to disentangle.  Fig.~\ref{fig4}(a) shows the intensity-dependent band-edge bleach dynamics of the defective QDs for four excitation intensities corresponding to different values of $N$. The dynamics for $N=0.21$ already exhibits a decay of $\approx$ 20$\%$, associated with carrier trapping on a $\approx$200 ps timescale. Over the same temporal window, however, the intensity-dependent dataset also displays the characteristic super-linear depletion associated with exciton-exciton annihilation.

The nonlinear-order decomposition by Malý~\textit{et al.} provides a direct way to constrain the kinetic model. As shown above, the 3$^{\mathrm{rd}}$-order component is exactly the single-exciton dynamics, $S^{(3)}(t)=d_1(t)$. Since exciton-exciton annihilation requires at least two excitons, it is absent from $S^{(3)}(t)$, which depends only on radiative recombination, carrier trapping, and the fraction $p_T$ of defective QDs. Fig.~\ref{fig4}(b) shows the experimentally extracted $S^{(3)}(t)$ (red circles), together with the fit obtained from the stochastic trapping model governed by Eq.~\ref{trapping_escape_rate} (black line). This fit yields $\kappa_R= (7\,\mathrm{ns})^{-1}$, $\kappa_{T}=(115\,\mathrm{ps})^{-1}$, and $p_T=0.17$. These parameters can then be fixed when analyzing the full intensity-dependent dataset in Fig.~\ref{fig4}(a) with a stochastic kinetic model that also includes exciton-exciton annihilation. For a QD containing $n$ excitons, with $n_T=1$ when a trap is available and $n_T=0$ otherwise, the total escape rate is
\begin{equation}
    \lambda_{(n,n_T)}
    =
    n\, \kappa_R
    +
    \binom{n}{2}\kappa_A
    +
    n\, n_T\, \kappa_{T}.
    \label{complete_escape_rate}
\end{equation}
With $\kappa_R$, $\kappa_{T}$, and $p_T$ fixed by the 3$^{\mathrm{rd}}$-order response, the annihilation rate $\kappa_A$ is the only remaining kinetic free parameter in the global fit of the intensity-dependent dynamics in Fig.~\ref{fig4}(a). The resulting fits are shown as black solid lines and yield $\kappa_A=(150\,\mathrm{ps})^{-1}$, which is comparable to $\kappa_{T}$. Such an overlap between trapping and multiexciton dynamics on the hundred-picosecond timescale is well documented in colloidal QDs~\cite{Smith2015, Sewall2008, Beane2016, Florio2025}. Here, nonlinear-order separation provides the critical additional constraint required to disentangle two kinetic processes that would otherwise remain strongly correlated in a direct global analysis. 

Having disentangled the trapping and annihilation rates, we can now evaluate their combined effect on $S^{(2k+1)}(t)$ and $d_n(t)$. Since the nonlinear signals due to each mechanism alone have opposite signs, see Fig.~\ref{fig2}(a) and \ref{fig3}(a), and develop on similar timescales, partial cancellation is expected. Fig.~\ref{fig4}(b) shows the signals extracted from the fit of the experimental data in Fig. \ref{fig4}(a). Their sign and temporal hierarchy closely resemble those obtained for the pairwise annihilation model in Fig.~\ref{fig2}: $S^{(5)}(t)$ is negative and the subsequent orders alternate in sign. Thus, taken alone, the nonlinear signals still appear predominantly annihilation-like and do not directly reveal the simultaneous presence of trap saturation. This is consistent with the intensity-dependent dynamics in Fig.~\ref{fig4}(a), which display the super-linear depletion characteristic of exciton-exciton annihilation. Therefore, an analysis only of $S^{(2k+1)}(t)$ could completely overlook the nonlinear effect of the traps.

The interplay between the two nonlinear mechanisms becomes clearer when the sub-ensemble dynamics $d_n(t)$ are considered, as shown in Fig. \ref{fig4}(c). At long times, a clear separation emerges between $d_1(t)$ and $d_{2-4}(t)$. The single-exciton trajectory, $d_1(t)=S^{(3)}(t)$, contains the fast trapping contribution together with the much slower radiative decay. Starting from $n=2$, the dynamics contain the combined effect of nonlinearities due to annihilation, as in Fig. \ref{fig2}(b), and saturable trapping in Fig. \ref{fig3}(b). If only annihilation were present, all $d_n(t)$ would collapse to the same population after its characteristic timescale. On the other hand, if only saturable trapping were present, each $d_n(t)$ could decrease  by at most one exciton on the trapping timescale. When both processes are present on 20\% of the particles and only annihilation is present in the remaining 80\%, $d_1(t)$ drops to around 0.8 following the trapping timescale. However, for $n \ge 2$, Fig. \ref{fig4}(c) shows that $d_n(t)$ never decreases below one exciton. This can be understood by considering that the trapping rate is slightly faster than the annihilation rate, so that we expect to have had at least one trapping event before annihilation is finished. Once the population reaches $n=1$, annihilation becomes inactive and further evolution is dictated by radiative recombination on its much longer timescale. Thus, while the $S^{(2k+1)}(t)$ closely resemble a case where only annihilation is present, the $d_n(t)$ immediately reveal the presence of a second nonlinearity of opposite sign. This case remains distinguishable from a model without saturable trapping, reported by control simulations in Supplementary Note~\ref{SI_no_saturation}. There, we see that removing trap saturation eliminates the $n$-dependent long-time offset and restores the common asymptotic behavior of the pure annihilation model. It is also interesting to note that, although the parameters of our sample do not produce a perfect cancellation of higher order $S^{(2k+1)}(t)$, such a set of parameters is physically accessible. In Supplementary Note~\ref{SI_cancellation} we show that increasing the defective fraction to $p_T=0.45$, the two effects become sufficiently similar in magnitude that the higher-order nonlinear signals are strongly suppressed by their opposite contributions, with only $S^{(5)}(t)$ remaining clearly present. Even in that case, inspection of $d_n(t)$ clearly reveals the presence of larger multiexciton populations exposed to nonlinear dynamics of opposite signs.

\section{Discussion and Conclusions}

The framework developed here establishes a natural distinction between two classes of nonlinear population dynamics: interaction-driven and saturation-driven. In both cases, the experimentally separated nonlinear signals provide a basis for reconstructing the excitation-number-resolved dynamics $d_n(t)$  without assuming a kinetic model. Interaction-driven nonlinearities result from a many-body kinetic process and exhibit a characteristic lowest order contribution $S^{(2k+1)}(t)$ determined by the minimum number of excitations required for that process to occur. For progressively larger initial populations, nonlinear signals of increasingly higher orders are required to correct for the changing number of combinations and permutations available as the population dynamically changes. Hence, the same pairwise interaction can generate $S^{(7)}(t)$, $S^{(9)}(t)$, and higher-order signals without requiring elementary three-, four-, or higher-body interactions. Since the statistic coefficients are fixed by the initial population, these signals exhibit progressively delayed rises, as they correct the signal only after previous interactions have taken place.

Saturation-driven nonlinearities follow a different logic. Here, the elementary kinetic event itself may involve only a single excitation, while the nonlinearity originates from competition for a finite-capacity state or channel. Once that capacity is exhausted, kinetic pathways that would have remained independent in the absence of saturation become mutually exclusive. The higher nonlinear orders then quantify competition among progressively larger numbers of excitations for the same limited set of final states. Thus, interaction-driven nonlinearities arise because the number of possible interactions changes with excitation number, whereas saturation-driven nonlinearities arise because the number of available final states does not increase accordingly.

The alternating signs observed in both classes can be understood from the same population-level perspective. The reconstruction of an $n$-excitation trajectory from progressively higher nonlinear orders involves successive corrections to the lower-order results. Each new term compensates for an over- or under-counting introduced by the preceding approximation, naturally producing alternating contributions in the examples considered here. Importantly, the alternating signs are not an automatic consequence of the $(-i/\hbar)^n$ prefactor found in the $n^{\mathrm{th}}$-order response function \cite{Hamm2011}. Indeed, Fig. \ref{fig3}(a) shows an example where $S^{(3)}$ and $S^{(5)}$ have the same sign, and this holds for any saturation-driven dynamic nonlinearity. When multiple nonlinearities of both classes are present, the specific parameters of the system determine which behavior dominates in $S^{(2k+1)}$.

Although the explicit transformation derived here follows from Poissonian preparation of independent excitations, the underlying construction is more general. For an arbitrary excitation-number distribution $P_N(n)$, the measured response can still be written as in Eq.~\ref{average_is_the_number}. Expanding $P_N(n)$ in powers of excitation intensity defines a corresponding linear transformation between the nonlinear-order components and the excitation-number-resolved dynamics. The Poissonian case considered here is therefore a particularly simple realization, for which this transformation admits the closed form derived above. Other excitation statistics would lead to different transformation coefficients, but the same general strategy can be applied whenever the excitation statistics are known, and the resulting mapping is invertible. We note, however, that deviations from the Poisson distribution are associated with effects like absorption saturation, which is not a dynamic nonlinearity and can be correctly accounted for with previous methods \cite{Maly2023, Rose2023, Krich2025, Luttig2026}.

In summary, we have established an invertible connection between experimentally accessible nonlinear pump--probe signals and sub-ensemble population dynamics for Poissonian excitations. Measurement of the nonlinear response up to $S^{(2n+1)}(t)$ is sufficient to reconstruct $d_n(t)$ algebraically, without specifying rate equations or a microscopic kinetic model. We have shown that $d_n(t)$ are not only more intuitive than $S^{(2n+1)}(t)$, but also provide direct insight into the many-body interactions and state-filling effects that may underlie the observed dynamics. It is also worth emphasizing that these results are not intrinsically restricted to semiconductor QDs. The essential requirement is not zero dimensionality itself, but the possibility of defining excitation-number-resolved sub-ensembles and a well-defined relation between excitation statistics and the measured nonlinear response. The framework can therefore be extended to other confined or mesoscopic photoexcited systems, including molecular aggregates, conjugated polymers, or other systems in which discrete excitation-number sectors can be meaningfully identified. In spatially extended systems, additional degrees of freedom such as exciton diffusion and spatial correlations may require a generalized state description, but the underlying separation between excitation statistics, excitation-resolved dynamics, and nonlinear-order observables remains applicable. Work in this direction is currently underway.

\begin{acknowledgments}
We thank Pavel Malý for useful discussions. This work has been supported under the National Recovery and Resilience Plan (NRRP), Mission 4 Component 2 Investment 1.4 – Call for tender No. 3138 of December 16, 2021 of the Italian Ministry of University and Research, funded by the European Union—NextGenerationEU [Award Number: CNMS named MOST, Concession Decree No. 1033 of June 17, 2022, adopted by the Italian Ministry of University and Research, CUP: B43C22000440001, Spoke 14 “Hydrogen and New Fuels”]. This research was supported by the Italian Ministry of Foreign Affairs – Directorate General for Cultural and Economic Promotion and Innovation and by the Israeli Ministry of Innovation, Science, and Technology (G.C. \& U.B., grant No. 1001585060). F.V., S.B. and F.V.A.C. acknowledge funding from the European Union (ERC, HOLOFAST, grant number 101117858). Views and opinions expressed are, however, those of the authors only and do not necessarily reflect those of the European Union or the European Research Council. Neither the European Union nor the granting authority can be held responsible for them. 
\end{acknowledgments}
\bibliography{apssamp}
\appendix
\onecolumngrid
\begin{center}
    {\Large\bfseries Supplementary Information}
\end{center}

\renewcommand{\appendixname}{}
\setcounter{section}{0}
\renewcommand{\thesection}{\arabic{section}}
\setcounter{figure}{0}
\renewcommand{\thefigure}{S\arabic{figure}}
\setcounter{equation}{0}
\renewcommand{\theequation}{S\arabic{equation}}

\section{Derivation of the connection between $d_n(t)$ and $S^{(2k+1)}(t)$}
\label{SI_derivation_dn_to_nls}
To connect the population dynamics $d_n(t)$ to the nonlinear-order decomposition, the ensemble response can be rewritten as a power series in the average number of absorbed photons per quantum dot, $N$. Under the excitation conditions considered here, $N$ is proportional to the pump fluence. Powers of $N$ therefore follow the same hierarchy as the pump-intensity powers used to separate the nonlinear components $S^{(2k+1)}(t)$.

Starting from the Poisson distribution in Eq.~\ref{poisson},
\begin{equation}
    P_N(n)=\frac{N^n}{n!}e^{-N},
\end{equation}
the exponential factor can be expanded as
\begin{equation}
    e^{-N}
    =
    \sum_{m=0}^{\infty}
    \frac{(-1)^m}{m!}N^m.
\end{equation}
Therefore,
\begin{equation}
    P_N(n)
    =
    \frac{N^n}{n!}
    \sum_{m=0}^{\infty}
    \frac{(-1)^m}{m!}N^m
    =
    \sum_{m=0}^{\infty}
    \frac{(-1)^m}{n!m!}N^{n+m}.
\end{equation}
Introducing $k=n+m$, such that $m=k-n$, gives
\begin{equation}
    P_N(n)
    =
    \sum_{k=n}^{\infty}
    \frac{(-1)^{k-n}}{n!(k-n)!}N^k.
    \label{Poisson_N_expansion}
\end{equation}

Substituting Eq.~\ref{Poisson_N_expansion} into
Eq.~\ref{average_is_the_number} and reordering the sums yields
\begin{equation}
    S(N,t)
    =
    \sum_{k=1}^{\infty}
    N^k
    \sum_{n=1}^{k}
    \frac{(-1)^{k-n}}{n!(k-n)!}d_n(t).
    \label{nlo_and_exciton_number_exp}
\end{equation}

Since $N$ is proportional to the pump intensity, Eq.~\ref{nlo_and_exciton_number_exp} can be compared order by order with the nonlinear intensity expansion in Eq.~\ref{Intensity_expansion}. This gives
\begin{equation}
    S^{(2k+1)}(t)
    \propto
    \sum_{n=1}^{k}
    \frac{(-1)^{k-n}}{n!(k-n)!}d_n(t).
    \label{nlo_signals_propto}
\end{equation}

In the present analysis, the fluence-dependent traces were converted into population-normalized ensemble responses by normalizing each trace to its maximum bleach amplitude and multiplying it by the corresponding value of $N$. With this normalization, the processed ensemble response is expressed on a scale proportional to the average exciton population. The proportionality constants in Eq.~\ref{nlo_signals_propto} are therefore absorbed into the definition of the extracted nonlinear components, yielding
\begin{equation}
    S^{(2k+1)}(t)
    =
    \sum_{n=1}^{k}
    \frac{(-1)^{k-n}}{n!(k-n)!}d_n(t).
    \label{nlo_signals}
\end{equation}

Equation~\ref{nlo_signals} defines a lower-triangular linear transformation from the excitation-number basis to the nonlinear-order basis, which can be inverted. Using the binomial inverse relation gives Eq.~\ref{dn_to_S(2k+1)},
\begin{equation}
    d_n(t)
    =
    \sum_{k=1}^{n}
    \binom{n}{k}
    k!\,
    S^{(2k+1)}(t).
    \label{exciton_number_signal_vs_nlo}
\end{equation}

\section{Derivation of the statistical decomposition of $d_n(t)$}
\label{SI_derivation_recursive_dn}

To connect the population dynamics $d_n(t)$ to the underlying kinetics, we start from the standard master-equation description of the excitation-number populations. For Markovian single-step kinetics, the ensemble of trajectories contributing to $d_n(t)$ can be partitioned according to the waiting time to the first kinetic event. This yields a recursive representation that separates trajectories that remain in the initial $n$-excitation state from those that undergo a transition and subsequently evolve from the state reached. Importantly, this form makes it possible to identify where many-body nonlinearities enter the population dynamics: either through the excitation-number dependence of the escape rate or through a change in the post-transition kinetics of the state reached.

Let $p_j^{(n)}(t)$ denote the probability that a system initially prepared with $n$ excitations contains $j$ excitations at time $t$. The corresponding excitation-number-resolved population dynamics is therefore
\begin{equation}
    d_n(t)=\sum_{j=0}^{n}j\,p_j^{(n)}(t).
\end{equation}
We first consider Markovian kinetics in which each kinetic event decreases the excitation number by one, $j\rightarrow j-1$, as schematically illustrated in Fig.~\ref{fig1}(d). In the simplest case, where the excitation number completely specifies the kinetic state, the state probabilities obey the master equation
\begin{equation}
    \frac{d p_j^{(n)}(t)}{dt}=\lambda_{j+1}p_{j+1}^{(n)}(t)-\lambda_j p_j^{(n)}(t),
\end{equation}
where $\lambda_j$ is the total escape rate from the $j$-excitation state.

For the initially occupied state, $j=n$, there is no incoming population. Therefore,
\begin{equation}
    \frac{d p_n^{(n)}(t)}{dt}=-\lambda_n p_n^{(n)}(t),
    \qquad
    p_n^{(n)}(0)=1,
\end{equation}
which gives
\begin{equation}
    p_n^{(n)}(t)=e^{-\lambda_n t}.
\end{equation}
Since the excitation number can only decrease, remaining in the $n$-excitation state is equivalent to having experienced no kinetic event. Thus, $e^{-\lambda_n t}$ is the probability that no event has occurred up to time $t$.

Let $T_n$ denote the waiting time to the first kinetic event. The $n$-excitation state survival probability is therefore
\begin{equation}
    \Pr(T_n>t)=e^{-\lambda_n t},
\end{equation}
and the corresponding waiting-time probability density is
\begin{equation}
    f_n(\tau)=\frac{d}{d\tau}[1-\Pr(T_n>\tau)]=-\frac{d}{d\tau}\Pr(T_n>\tau)
    =\lambda_n e^{-\lambda_n\tau}.
\end{equation}
Hence,
\begin{equation}
    f_n(\tau)\,d\tau=\lambda_n e^{-\lambda_n\tau}\,d\tau
\end{equation}
is the probability that the first kinetic event occurs between $\tau$ and $\tau+d\tau$.

The average population at time $t$ can now be separated according to whether the first event has occurred before $t$. If no event has occurred, the system still contains all $n$ initial excitations and contributes
\begin{equation}
    n\,\Pr(T_n>t)=n\,e^{-\lambda_n t}
\end{equation}
to the ensemble-averaged population. Alternatively, if the first event occurs at time $\tau<t$, the system reaches a state with $n-1$ excitations and then evolves for the remaining time $t-\tau$. Denoting the conditional population dynamics following this first event by $\widetilde d_{n-1}(t-\tau)$, the contribution of trajectories whose first event occurs between $\tau$ and $\tau+d\tau$ is \begin{equation}
    f_n(\tau)\,\widetilde d_{n-1}(t-\tau)\,d\tau.
\end{equation}
Averaging over all possible first-event times therefore gives Eq.~\ref{convolution}
\begin{equation}
    d_n(t)=n\,e^{-\lambda_n t}+\int_0^t\lambda_n e^{-\lambda_n\tau}
    \widetilde d_{n-1}(t-\tau)\,d\tau.
\end{equation}
Eq.~\ref{convolution} thus partitions the ensemble into two mutually exclusive classes of trajectories: those that have not yet undergone any kinetic event at time $t$, and those whose first event occurred at some earlier time $\tau$. Importantly, $\widetilde d_{n-1}(t)$ denotes the dynamics of the state reached after the first transition and need not, in general, be identical to $d_{n-1}(t)$, which describes a system initially prepared with $n-1$ excitations. If the excitation number completely specifies the kinetic state, then $\widetilde d_{n-1}(t)=d_{n-1}(t)$. If, instead, the first event also modifies an additional kinetic variable, such as the occupancy of a finite-capacity trapping site, the subsequent dynamics can differ even though the excitation number is the same, and $\widetilde d_{n-1}(t)\neq d_{n-1}(t)$.

The same first-event decomposition can be extended to several competing kinetic channels or to transitions with different changes in excitation number by introducing the corresponding event-specific rates and conditional post-transition dynamics.

\section{Validation of the kinetic model of Section \ref{sec:interaction-driven_nl}}
\label{SI_experimental_data}

\begin{figure}[hbtp]
    \centering
    \includegraphics[scale=1]{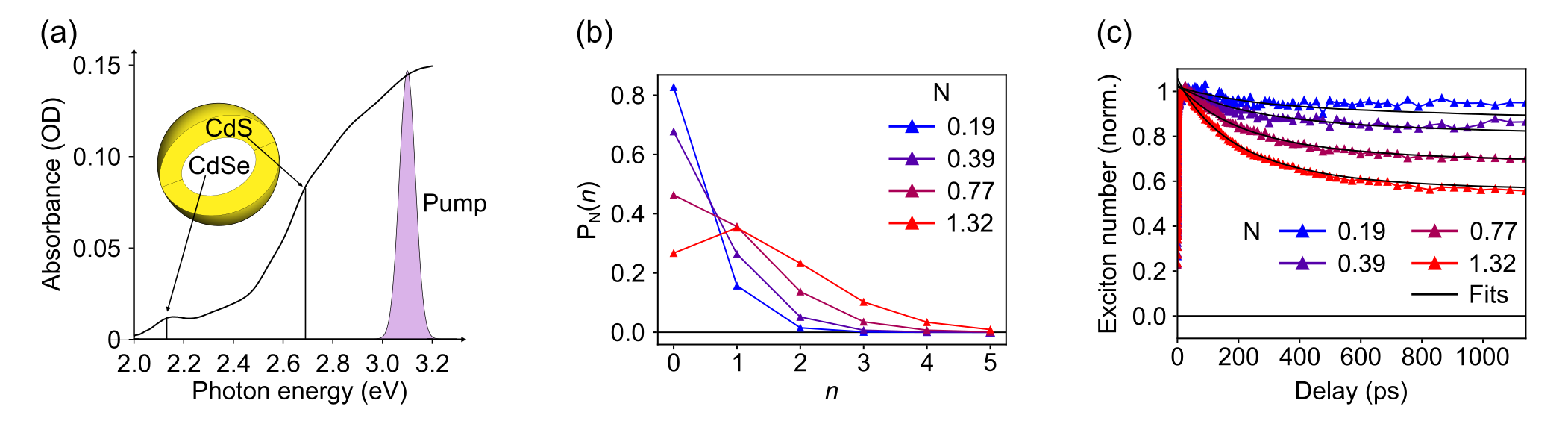}
    \caption{(a) Ensemble absorption spectrum of the QDs together with the spectrum of the 3.10 eV excitation pulses used in the ultrafast experiments. (b)  Initial excitation statistics of QDs in panel a. (c) Transient absorption dynamics of the band-edge states following photoexcitation at 3.1 eV, normalized to the average number of excitons per QD, \textit{N}. The fits are shown as black solid lines.}
    \label{fig:exp_data}
\end{figure}

The kinetic model defined by the escape rate in Eq.~\ref{pairwise_escape_rate} of the main text was previously validated against experimental data from core--shell CdSe/CdS quantum dots~\cite{Florio2025}. Figure~\ref{fig:exp_data} summarizes the experimental conditions and the comparison between the measured dynamics and the model.

Figure~\ref{fig:exp_data}(a) shows the linear absorption spectrum of the QDs dispersed in toluene, together with the pump spectrum in purple. The band-edge transition associated with the CdSe core lies at $\approx2.1$ eV.  The pump photon energy is centered at $3.1$~eV, providing an excess energy of $\approx1$ eV to the photoexcited electron--hole pairs. At this excitation energy, absorption occurs in a spectral region with a high density of available states, such that successive absorption events can be reasonably treated as statistically independent. Accordingly, the initial excitation-number distribution is well described by Poisson statistics (Eq.~\ref{poisson} of the main text). Figure~\ref{fig:exp_data}(b) shows the corresponding Poisson distributions for the four excitation conditions used experimentally, characterized by different values of the average exciton number $N$. 

Figure~\ref{fig:exp_data}(c) shows the normalized transient band-edge bleach dynamics measured for the same four excitation conditions. The results of the stochastic kinetic model are shown as black solid lines. The radiative and exciton-exciton annihilation rates were found to be $\kappa_R=(29.6\,\mathrm{ns})^{-1}$, as determined from complementary TRPL measurements, and $\kappa_A=(250\,\mathrm{ps})^{-1}$, respectively. The model accurately reproduces the progressively faster population depletion observed with increasing $N$. This behavior originates from the increasing fraction of QDs initially prepared with multiexciton populations, for which pairwise annihilation provides an additional population-loss channel that is approximately two orders of magnitude faster than radiative recombination. In the main text, $\kappa_A=(150\,\mathrm{ps})^{-1}$ is used to better highlight the effect of exciton-exciton annihilation on the decay in the 1 ns time window.

\section{Control simulations without trap saturation}
\label{SI_no_saturation}

\begin{figure}[hbtp]
    \centering
    \includegraphics[scale=1]{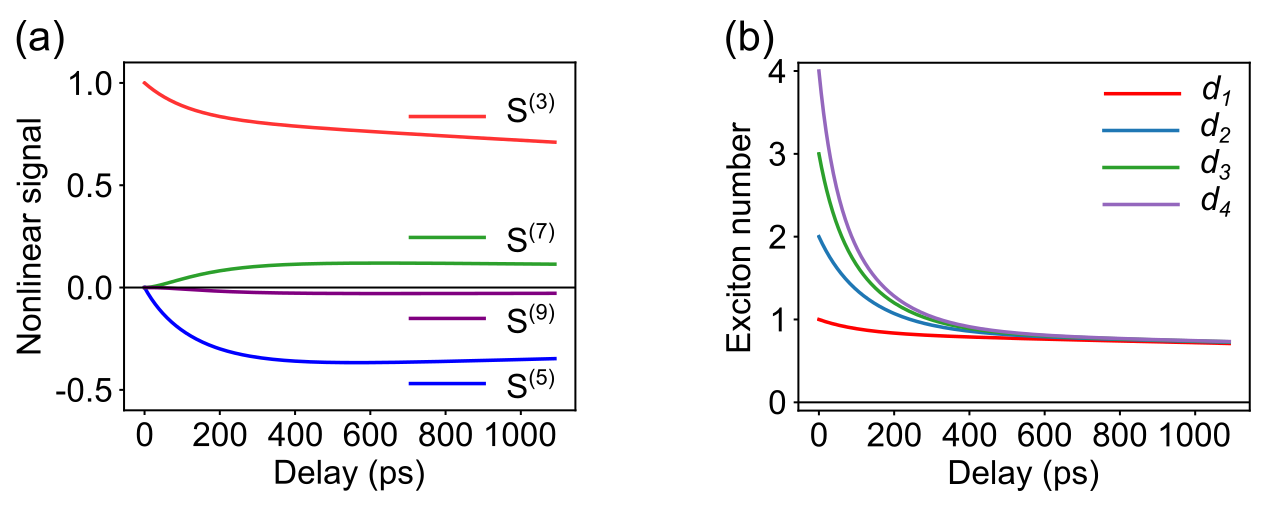}
    \caption{(a) 3$^{\mathrm{rd}}$-, 5$^{\mathrm{th}}$-, 7$^{\mathrm{th}}$-, and 9$^{\mathrm{th}}$-order signals for an artificial ensemble of defective QDs using the same kinetic parameters extracted from the experimental data in Fig.~\ref{fig4}, but assuming non-saturable carrier trapping. (b) Dynamics for a sub-ensemble of particles with 1 to 4 excitons at $t=0$.}
    \label{fig:no_saturation}
\end{figure}

Here we consider a trapping model where no saturation is present, in order to assess which different signatures would emerge both in $S^{(2k+1)}$ and $d_n$ in contrast to the saturable trap model. Specifically, we assume that a large density of trap states is present such that trap states are always available to all excitons. All kinetic parameters are kept identical to those extracted from the experimental data in Fig.~\ref{fig4}. Figure~\ref{fig:no_saturation}(a) shows the resulting nonlinear components, which remain qualitatively very similar to their saturable-trapping counterparts in Fig.~\ref{fig4}(b). The reconstructed sub-ensemble dynamics $d_n(t)$, however, reveal a clear difference. In the absence of trap saturation, the trajectories initialized with different exciton numbers converge to the same population dynamics once the exciton-exciton annihilation is complete, reproducing the behavior observed for the pure annihilation model in Fig.~\ref{fig2}(b). This contrasts with Fig.~\ref{fig4}(c), where a persistent separation between $d_1(t)$ and $d_{2-4}(t)$ is observed. The origin of this difference is that excitons no longer compete for a finite-capacity trapping channel, so that carrier trapping remains a linear single-exciton process and, by itself, generates only the third-order response. The higher-order nonlinearities are therefore generated by exciton-exciton annihilation, whereas the excitation-number-dependent long-time offset observed in Fig.~\ref{fig4}(c) is a direct consequence of trap saturation.

\section{Near-complete compensation of opposite dynamic nonlinearities}
\label{SI_cancellation}

\begin{figure}[hbtp]
    \centering
    \includegraphics[scale=1]{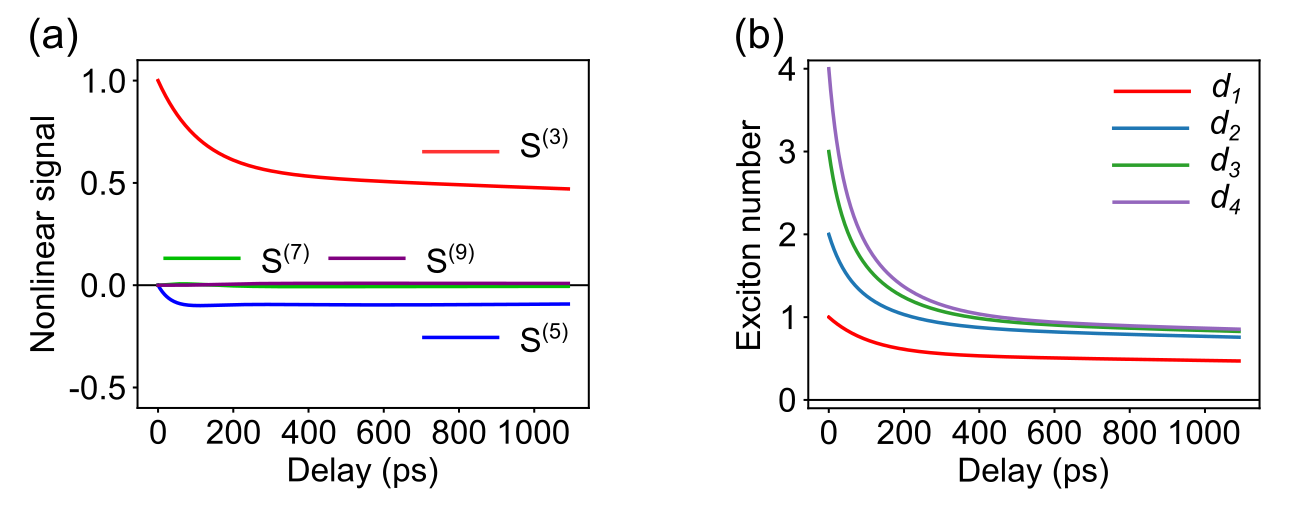}
    \caption{(a) 3$^{\mathrm{rd}}$-, 5$^{\mathrm{th}}$-, 7$^{\mathrm{th}}$-, and 9$^{\mathrm{th}}$-order signals for an artificial ensemble of defective QDs with $p_d=0.45$. (b) Dynamics for a sub-ensemble of particles with 1 to 4 excitons at $t=0$.}
    \label{fig:cancell}
\end{figure}

The experimental data and order-selective analysis in Fig.~\ref{fig4} show that kinetic processes producing opposite nonlinear intensity dependencies can partially compensate within the same nonlinear orders. Here, we illustrate that this compensation can become nearly complete. Keeping all experimentally extracted kinetic parameters fixed, we increase only the defective fraction to $p_T=0.45$. Figure~\ref{fig:cancell}(a) shows the resulting nonlinear components. The third-order response contains the radiative-recombination and carrier-trapping dynamics, whereas the fifth-order response contains the exciton-exciton annihilation contribution but is strongly reduced by the opposing fifth-order contribution generated by trap saturation. For the seventh- and ninth-order components, this compensation becomes even stronger, suppressing the net nonlinear signals close to the noise-floor level. 

The sub-ensemble dynamics $d_n(t)$, reconstructed from these nonlinear components through Eq.~\ref{dn_to_S(2k+1)} of the main text, are shown in Fig.~\ref{fig:cancell}(b). Despite the weak higher-order signals, the reconstructed $d_n(t)$ retain clear signatures of both nonlinear mechanisms. Exciton-exciton annihilation produces progressively faster depletion as the initial exciton number increases, while trap saturation appears through the $n$-dependent asymptotic offset, which is even more pronounced here than in Fig.~\ref{fig4}(c) of the main text. This example therefore shows that weak higher-order nonlinear signals do not necessarily imply negligible multiexciton dynamics, but can instead result from cancellation between nonlinear processes with opposite signs.
\end{document}